\documentstyle[aps,epsf,preprint,eqsecnum]{revtex}
\begin{document}
 \makeatletter
 \newdimen\ex@
 \ex@.2326ex
 \def\dddot#1{{\mathop{#1}\limits^{\vbox to-1.4\ex@{\kern-\tw@\ex@
  \hbox{\tenrm...}\vss}}}}
 \makeatother
\thispagestyle{empty}
{\baselineskip0pt
\leftline{\large\baselineskip16pt\sl\vbox to0pt{\hbox{\it Department of Physics}
               \hbox{\it Osaka City  University}\vss}}
\rightline{\large\baselineskip16pt\rm\vbox to20pt{\hbox{OCU-PHYS-191}
            \hbox{AP-GR-8}
            \hbox{\today} 
\vss}}%
}
\vskip3cm
\begin{center}{\large 
{\bf How Does Naked Singularity Look?}
}
\end{center}
\begin{center}
 {\large Ken-ichi Nakao\footnote{e-mail: knakao@sci.osaka-cu.ac.jp}, 
Naoki Kobayashi 
and Hideki Ishihara\footnote{e-mail: ishihara@sci.osaka-cu.ac.jp}
} \\
{\em Department of Physics, Graduate School of Science,
~Osaka City University} \\
{\em Osaka 558-8585,~Japan}
\end{center}
\begin{abstract}
There are non-radial null geodesics 
emanating from the shell focusing singularity formed at the 
symmetric center in a spherically symmetric dust collapse. 
In this article, assuming the self-similarity in the region 
filled with the dust fluid, we study these singular null 
geodesics in detail. We see the time evolution 
of the angular diameter of the central naked singularity and show that 
it might be bounded above by the value 
corresponding to the circular null geodesic in 
the Schwarzschild spacetime. 
We also investigate the angular frequency of a physical field 
which propagates along the singular null geodesic and 
find that it depends on the impact parameter. 
Further, we comment on the non-uniformity of the 
topology of the central naked singularity. 
\vskip0.5cm
\noindent
PACS number(s): 04.20.Dw, 04.20.Gz
\end{abstract}

\newpage
\section{Introduction}

Gravitational collapse is one of the most important issues in general  
relativity and in the theoretical research for this phenomenon, 
the Lema\^{\i}tre-Tolman-Bondi(LTB) solution which describes 
the motion of spherically symmetric dust fluid has played an important 
role. This is a very simple analytic solution 
of Einstein equations but has substantial physical contents. 
The simplest but physically significant member of this solution is 
the case of a homogeneous dust sphere, which 
is called the Oppenheimer-Snyder solution. 
This solution is the first example to give an elegant picture 
of a black-hole formation and  
led to the so-called cosmic censorship conjecture; 
roughly speaking, this conjecture states that 
spacetime singularities except for the Big Bang are 
not visible for any observer if it is resulted from physically 
reasonable initial conditions\cite{ref:penrose69}. 
However, this conjecture has not yet been proven although it 
plays very crucial roles in the proof of the singularity 
theorems\cite{ref:penrose65,ref:hawking67,ref:hp70} 
and the area theorem of black hole\cite{ref:hawking72}. 
Rather LTB solution seems to be a candidate of the counterexample 
of this conjecture. 
Eardley and Smarr pointed out that a shell focusing 
singularity formed at the symmetric center in an {\it inhomogeneous}  
dust sphere can be naked, i.e., visible 
in principle\cite{ref:ES79}. After this discovery, several 
theoretical efforts revealed the genericity of the shell 
focusing naked singularity formation in the LTB 
solution\cite{ref:christodoulou84,ref:Newman86,ref:JD92,ref:JD93,ref:SJ96,ref:JJS96}; there are radial null geodesics emanating from the central shell
focusing singularity. 

The central naked singularity of the dust sphere 
seems to be a ``point'' since it forms from an infinitesimal 
portion at the symmetric center of the dust sphere. 
However, it is not true. Recently, Mena and Nolan 
showed that there are non-radial 
null geodesics emanating from this central naked 
singularity\cite{ref:MN01}, where non-radial means that 
it has a non-vanishing angular momentum. 
Deshingkar, Joshi and Dwivedi (DJD) studied numerically the trajectories 
of those geodesics within the dust fluid in detail\cite{ref:DJD02}. 
More recently, Nolan and Mena studied various 
invariant quantities associated with this central naked singularity and 
further discussed the topology of this naked singularity\cite{ref:NM02}. 

It is the main purpose of this article to reveal how the central 
naked singularity is observed for distant observers 
outside the dust sphere. For this purpose, we investigate both 
the radial and non-radial null geodesics emanating from the 
central naked singularity in the self-similar dust sphere. 
Assuming that the dust sphere has a finite mass, 
we investigate the angular diameter of the naked singularity, or
equivalently, the largest impact parameter of the null 
geodesics from the naked singularity; we show its time evolution. 
We also clarify the non-uniformity of topology of the central 
naked singularity for the self-similar case analyzed  
by Nolan and Mena\cite{ref:NM02}. 

We also study angular frequencies of physical fields which propagate 
along the null geodesics emanating from the central naked singularity. 
DJD showed that the redshift between 
the central naked singularity and any off-center observer is 
infinite\cite{ref:DJD02}. Hence even if physical fields with finite 
frequencies are excited at the naked singularity,  
``physical'' information of the naked 
singularity can not reach distant observers. 
However, it would not be natural that physical fields 
with {\it finite} frequencies 
are generated at the naked singularity where the energy 
density of the dust fluid and spacetime curvature diverge. 
In reality, linear perturbation analyses of LTB solution 
show that the spacetime curvature associated to 
gravitational-wave perturbations indefinitely grows in the limit to 
the Cauchy horizon\cite{ref:IHN00,ref:NIH01}. 
This means that the frequencies of the linear gravitational waves 
generated at the naked singularity will be infinite even in purely 
classical dynamics. 
Physical fields excited by quantum effects also have infinite 
frequencies at the naked singularity\cite{ref:BSVW98a,ref:BSVW98b,ref:VW98,ref:HIN00a,ref:HIN00b,ref:HINDTV01}. 
Hence, in this article, we study the angular frequencies of 
the physical fields emanating from the central naked singularity 
with appropriately {\it infinite} frequencies. 

This paper is organized as follows. In Sec.II, we present 
the model of background spacetime, which describes the gravitational 
collapse of a self-similar dust sphere with a finite mass.  
In Sec.III, we derive asymptotic behaviors of the null geodesics near 
the naked singularity and 
the basic equations for numerical calculations to see the intermediate 
behaviors of the null geodesics inside the dust sphere. 
In this section, we comment on the previous study about the topology 
of the naked singularity by Nolan and Mena. 
In Sec.IV, we explain the null geodesics in the exterior vacuum region. 
In Sec.V, we present the numerical results for the apparent size of the 
naked singularity, i.e., the angular diameter of the central naked 
singularity, or equivalently, the largest impact parameter 
of the singular null geodesic.  
In Sec.VI, the frequencies of physical fields propagating along 
null geodesics emanating from the central naked singularity 
are investigated. Finally, Sec.VII is devoted for summary and discussion. 

In this article, we adopt the unit $G=c=\hbar=1$ and 
follow the sign conventions of the metric and 
Riemann tensors in Ref.\cite{ref:wald}. 

\section{Collapsing Dust Sphere}

In this section, we show the model of background spacetime; the 
gravitational collapse of a self-similar dust sphere with a finite 
mass. We assume that the dust fluid exists only within a 
finite comoving region $0\leq r \leq r_{\rm s}$, 
where $r_{\rm s}$ is a positive constant. 

\subsection{Interior Solution}

Dust fluid elements move along timelike geodesics. We can adopt  
the tangent vector of these geodesics as a temporal coordinate basis. 
Then the line element is written in the form
\begin{equation}
ds^{2}=-dt^{2}+{{R'}^{2}(t,r)dr^{2}\over 1+f(r)}+R^{2}(t,r)(d\theta^{2}
+\sin^{2}\theta d\varphi^{2}),
\end{equation}
where a prime denotes the derivative with respect to the comoving radial 
coordinate $r$, and $f(r)$ is an arbitrary function. 
In this coordinate system, the components of the 
4-velocity of the dust fluid elements are $u^{\mu}=\delta_{t}^{\mu}$.  
Einstein equations and the equation of motion for the dust fluid 
lead to
\begin{eqnarray}
{\dot R}^{2}&=&{F(r)\over R}+f(r), \label{eq:energy-eq}\\
\rho(t,r)&=&{F'\over 8\pi R'R^{2}}, \label{eq:rho-sol}
\end{eqnarray}
where a dot means the temporal derivative, 
$\rho(t,r)$ is the rest mass density of the dust fluid, and 
$F(r)$ is an arbitrary function. From Eq.(\ref{eq:rho-sol}), 
the arbitrary function $F(r)$ is regarded as the twice of a mass within the 
comoving radius $r$. Then Eq.(\ref{eq:energy-eq}) 
is regarded as an energy equation for a dust fluid element labeled by
$r$ and hence $f(r)$ corresponds to the specific energy of it. 

For simplicity, hereafter we assume the self-similarity 
for this spacetime; 
all the dimensionless variables are the functions of 
the self-similar variable $x:=t/r$ only. 
Then the arbitrary function $F(r)$ is written in the form
\begin{equation}
F(r)=\Lambda_{0}r, \label{eq:mass-eq}
\end{equation}
where $\Lambda_{0}$ is a positive dimensionless constant. 
The specific energy $f(r)$ should vanish by this self-similar 
hypothesis.  
Then the solution of Eq.(\ref{eq:energy-eq}) is given in the form
\begin{equation}
R=r\left(1-{3\sqrt{\Lambda_{0}}\over 2}x\right)^{2\over3}, 
\label{eq:R-solution}
\end{equation}
where using a freedom to re-scale the radial coordinate $r$, we 
have chosen an arbitrary function associated to the 
temporal integration so that $R=r$ at $t=0$. 
The dust fluid element at $r$ forms the shell focusing singularity 
at $t=2r/3\sqrt{\Lambda_{0}}$. 

\subsection{Exterior Solution}

The exterior region is assumed to be vacuum and hence is described by the 
Schwarzschild geometry by Birkhoff's theorem. 
Using the Schwarzschild static chart, the line element of 
the exterior region is written as
\begin{equation}
ds^{2}=-C(R)dT^{2}+{dR^{2}\over C(R)}+R^{2}\left(d\theta^{2}
+\sin^{2}\theta d\varphi^{2}\right),
\end{equation}
where the function $C(R)$ is defined by 
\begin{equation}
C(R):=1-{2M\over R}:=1-{\Lambda_{0}r_{\rm s}\over R}.
\end{equation}
The angular coordinates $\theta$ and $\varphi$ are common to both 
the interior and exterior regions. 

The trajectory of the surface $r=r_{\rm s}$ is a marginally 
bound timelike geodesic in the Schwarzschild spacetime by continuity. 
Thus the integration of the temporal 
component of geodesic equations and the normalization 
condition of the 4-velocity vector lead to 
\begin{eqnarray}
{dT\over dt}&=&{1\over C(R)},  \label{eq:u0-sol} \\
{dR\over dt}&=&-\sqrt{2M\over R}, \label{eq:u1-sol}
\end{eqnarray}
where $t$ is the proper time which agrees with the interior 
temporal coordinate, and to derive the second equation, we have used 
the fact that the dust sphere is collapsing. 

From Eqs.(\ref{eq:u0-sol}) and (\ref{eq:u1-sol}), 
the trajectory of the surface is determined by
\begin{equation}
{dR\over dT}=-C(R)\sqrt{2M\over R}.
\end{equation}
The above equation is easily integrated and we obtain
\begin{equation} 
T-T_{\rm d}=-{2\over 3}\sqrt{R\over 2M}\left(R+6M\right)
-2M\ln\Biggl|{\sqrt{R}-\sqrt{2M}\over \sqrt{R}+\sqrt{2M}}\Biggr|,
\label{eq:DS-trajectory}
\end{equation}
where $T_{\rm d}$ is an integration constant.

\section{Null Geodesics inside the Dust Sphere}

By virtue of the spherical symmetry, it does not loose any generality 
to assume that the null geodesic is confined within the $\theta=\pi/2$ 
plane and hence we do so. 
Then the tangent vector $k^{\mu}$ of the future directed  
outgoing null geodesic in the LTB spacetime 
is expressed in the form
\begin{eqnarray}
k^{t}&=&{dt\over d\lambda}=:{{\cal P}\over R}, \\
k^{r}&=&{dr\over d\lambda}={\sqrt{{\cal P}^{2}-l^{2}}\over RR'},\\
k^{\theta}&=&{d\theta\over d\lambda}=0,\\
k^{\varphi}&=&{d\varphi\over d\lambda}
={l\over R^{2}},
\label{eq:kr-def}
\end{eqnarray}
where $\lambda$ is the affine parameter 
and $l$ is an integration constant which corresponds to 
the conserved angular momentum. Without loss of 
generality, we can assume $l\geq0$. 
We call a null geodesic of $l=0$ the {\it radial} one while 
a null geodesic of $l>0$ is called the {\it non-radial} one. 
${\cal P}$ should be larger than or equal to $l$ so that 
$k^{t}$ is positive and $k^{r}$ is real. 
In this article, we are interested in the null geodesics emanating from the 
central naked singularity; we call those the {\it singular null geodesics}. 

Instead of the affine parameter 
$\lambda$, we adopt the comoving radius $r$ to parameterize 
the null geodesic. 
Then the geodesic equations become  
\begin{eqnarray}
r{d{\cal P}\over d r}&=&
-{1\over W^{4}}\sqrt{\Lambda_{0}({\cal P}^{2}-l^{2})}
~r^{2(1-\alpha)}
+{{\cal P}\over3W^{3}}\left(W^{3}+2
r^{3(1-\alpha)/2}\right). \label{eq:P-equation}\\
{dR\over dr^{\alpha}}&=&
{1\over3\alpha W^{2}}\left(W-{\Lambda_{0}^{1/2}{\cal P}
\over \sqrt{{\cal P}^{2}-l^{2}}}~r^{(1-\alpha)/2}\right)
\left(W^{3}+2r^{3(1-\alpha)/2}\right),
\label{eq:R-equation}
\end{eqnarray}
where we have introduced a new variable defined by
\begin{equation}
W:=\sqrt{R\over r^{\alpha}},
\end{equation}
and $\alpha$ is some positive constant which will be fixed 
in the following discussion. 

\subsection{Asymptotic Behavior near the Singularity}

As discussed by Joshi and Dwivedi\cite{ref:JD93}, singular null geodesics 
should behave near the singularity as
\begin{equation}
R\longrightarrow W_{0}{}^{2}r^{\alpha}~~~~~~{\rm for}~~r\longrightarrow0, 
\label{eq:R-asymptotic}
\end{equation}
where $W_{0}$ is some positive constant.  Assuming this behavior, 
we search for asymptotic solutions of both the radial and non-radial 
singular null geodesics in the neighborhood of the central 
naked singularity. 

\subsubsection{Radial Singular Null Geodesics}

The asymptotic behavior of the radial singular null 
geodesic have already been studied well by several 
researchers\cite{ref:JD92,ref:JD93,ref:SJ96,ref:JJS96}. 
However, we would like to again see it in detail since 
the knowledge about the radial singular null geodesics 
is useful for the analysis of non-radial singular null geodesics. 

For radial null geodesics, 
Eqs.(\ref{eq:R-equation}) and (\ref{eq:R-asymptotic}) lead to
\begin{equation}
W_{0}{}^{2}=\lim_{r\rightarrow0}{1\over3\alpha W_{0}{}^{2}}
\left(W_{0}-\Lambda_{0}^{1/2}r^{(1-\alpha)/2}\right)
\left(W_{0}{}^{3}+2r^{3(1-\alpha)/2}\right).
\label{eq:W0-eq-1}
\end{equation}
The above equation determines $W_{0}$ and $\alpha$. 
In order that the above equation has a positive root for $W_{0}$, 
$\alpha$ should be $1/3$ or unity. The solutions of $\alpha=1/3$ 
correspond to null geodesics from the origin before the formation of 
central singularity while the solutions of $\alpha=1$ are of our 
interest, i.e., the singular null geodesics. 
To see this fact, we rewrite Eq.(\ref{eq:R-solution}) in the form
\begin{equation}
t={2\over 3\Lambda_{0}^{1/2}}
\left\{1-\left({R\over r}\right)^{3/2}\right\}r.
\end{equation}
Using the above equation and Eq.(\ref{eq:R-asymptotic}), we find that 
$t\rightarrow-2W_{0}^{3}/3\Lambda_{0}^{1/2}$ for $r\rightarrow0$ 
along the null geodesic of $\alpha=1/3$. 
This means that the null geodesics of $\alpha=1/3$ start from the 
regular center $r=0$ at $t<0$ and hence these are not singular 
null geodesics. On the other hand, in the case of $\alpha=1$, 
$t$ approaches to zero for $r\rightarrow0$ along the null geodesic. 
Hence the null geodesics of $\alpha=1$ emanate from the central 
singularity $r=0$ at $t=0$. 

\begin{figure}
        \centerline{\epsfxsize 10cm \epsfysize 7cm \epsfbox{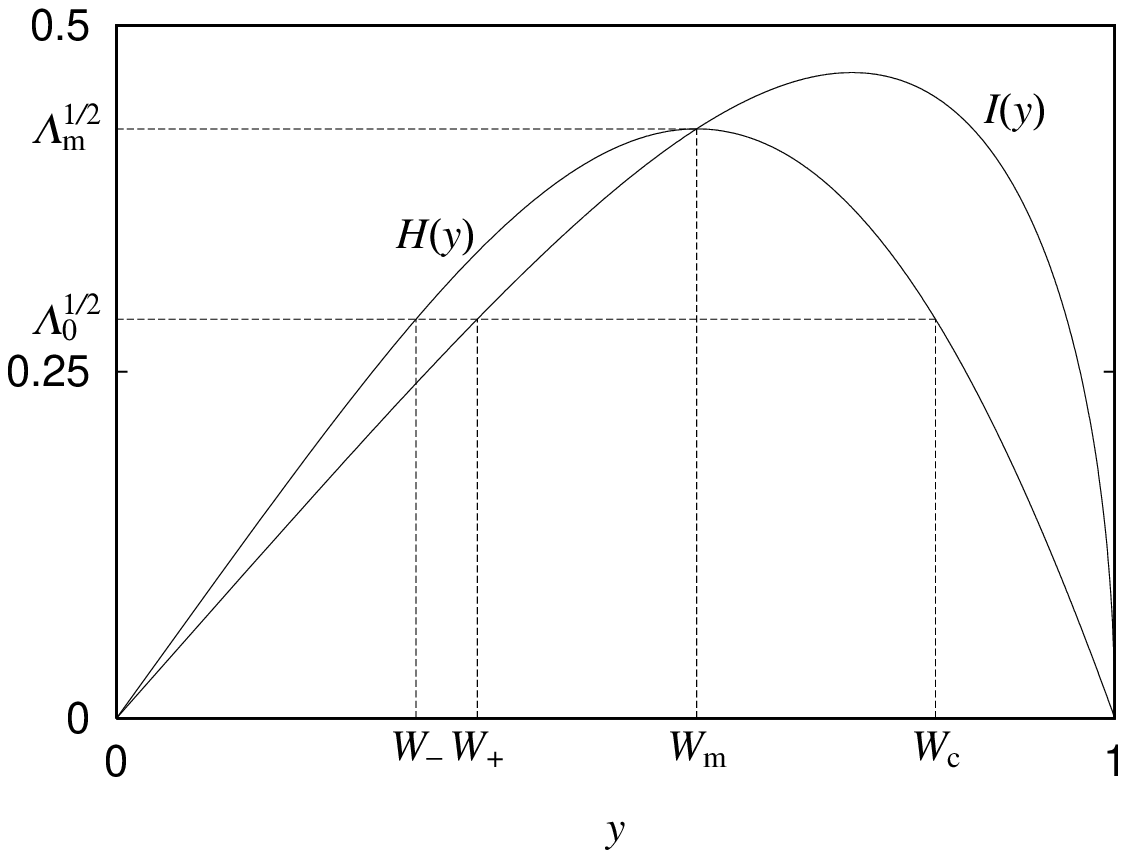}}
        \caption{The functions $H(y)$ and $I(y)$ are depicted. 
The roots of Eqs.(\ref{eq:W0-eq-1}) and (\ref{eq:W0-eq-2}) are also 
depicted for a given $\Lambda_{0}$ which is only one free parameter 
of the self-similar dust sphere. 
} 
\label{fg:H-I}
\end{figure}

Choosing $\alpha=1$, Eq.(\ref{eq:W0-eq-1}) leads to 
\begin{equation}
\Lambda_{0}^{1/2}=H(W_{0}):
={2W_{0}(1-W_{0}{}^{3})\over W_{0}{}^{3}+2}.
\label{eq:W0-eq-2}
\end{equation}
The radial singular null geodesics have to satisfy the above equation. 
Since $\Lambda_{0}^{1/2}$ is positive, $0<W_{0}{}<1$ should hold. 
We can easily find that there 
is a maximum of $H(W_{0})$ at 
\begin{equation}
W_{0}=W_{\rm m}:=\left(-5+3\sqrt{3}\right)^{1/3}.
\end{equation} 
Hence in order that a singular null geodesic exists, 
$\Lambda_{0}$ should be less than or equal to 
a critical value $\Lambda_{\rm m}$ defined by
\begin{equation}
\Lambda_{\rm m}:=H^{2}(W_{\rm m})
=\left\{4\left(26-15\sqrt{3}\right)\right\}^{2/3}.
\end{equation}
We depict $H(y)$ in Fig.\ref{fg:H-I}. 
In this article, we will focus on the case in which $\Lambda_{0}$ 
is smaller than the critical value $\Lambda_{\rm m}$. Then form  
Fig.\ref{fg:H-I}, we can easily see that for a given value 
of $\Lambda_{0}$, there are two positive real roots of 
Eq.(\ref{eq:W0-eq-2}). We denote the 
smaller root by $W_{-}$ and the other by $W_{\rm c}$. 

From Eq.(\ref{eq:P-equation}), we find
\begin{equation}
r{d{\cal P}\over dr}\longrightarrow
-\beta(W_{0}){\cal P}~~~~~
{\rm for}~~r\longrightarrow0,
\end{equation}
where
\begin{equation}
\beta(W_{0})
:={2-10W_{0}{}^{3}-W_{0}{}^{6}\over 3W_{0}{}^{3}(W_{0}{}^{3}+2)}.
\end{equation}
Since $\beta(W_{0})\neq0$ for $0<\Lambda_{0}<\Lambda_{\rm m}$, 
we obtain the asymptotic solution 
of ${\cal P}$ as
\begin{equation}
{\cal P}\longrightarrow {\rm Const.}\times r^{-\beta}. 
\label{eq:P-asymptotic}
\end{equation}

From the asymptotic behaviors of $W$ and $\cal P$, it is better 
to introduce the following variables, 
\begin{eqnarray}
P&:=&r^{\beta}{\cal P}, \\
\Delta W&:=&W-W_{0}.
\end{eqnarray}
Since $\Delta W$ will vanish and 
$P$ approaches to some finite positive value $P_{0}$, for 
$r\rightarrow 0$, Eqs.(\ref{eq:P-equation}) and (\ref{eq:R-equation}) lead to
\begin{eqnarray}
{d\Delta W\over d\ln r}&\longrightarrow& \beta \Delta W, \\
{dP\over d\ln r}&\longrightarrow&
{2P_{0}(2-5W_{0}^{3})\over W_{0}^{4}(W_{0}^{3}+2)}~\Delta W.
\end{eqnarray}
Integrating the above equations, we obtain asymptotic solutions 
for $r\rightarrow0$ as
\begin{eqnarray}
W&\longrightarrow& W_{0}+W_{*0}r^{\beta}, \label{eq:W-asymptotic-1}\\
{\cal P}&\longrightarrow& {P_{0}\over r^{\beta}}
\left\{1+{6(2-5W_{0}^{3})\over W_{0}
(2-10W_{0}^{3}-W_{0}^{6})}~W_{*0}r^{\beta} \right\}, 
\label{eq:P-asymptotic-1}
\end{eqnarray}
where $W_{*0}$ is an integration constant. 

Here again note that $W$ has to approach to $W_{0}$ 
in the limit of $r\rightarrow0$. This means that 
$\beta$ should be positive as long as $W_{*0}$ does not vanish. 
The positivity of $\beta$ leads to the following inequality 
\begin{equation}
0<W_{0}<W_{\rm m}. \label{eq:W0-inequality}
\end{equation}
Hence for non-vanishing $W_{*0}$, the smaller root $W_{0}=W_{-}$ 
of Eq.(\ref{eq:W0-eq-2}) has to be chosen (see Fig.1). 

The other root $W_{\rm c}$ of Eq.(\ref{eq:W0-eq-2}) 
is larger than $W_{\rm m}$ and hence $\beta(W_{\rm c})$ 
is negative. For such singular null geodesics, $W_{*0}$ has to 
vanish so that $W$ approaches to $W_{\rm c}$ for $r\rightarrow0$. 
This implies that $W$ should be always equal to 
$W_{\rm c}$ by Eq.(\ref{eq:R-equation}). 
This null geodesic is a generator of the Cauchy horizon. 
Substituting $\alpha=1$, $l=0$ and $W=W_{\rm c}$ into 
Eq.(\ref{eq:P-equation}), we can easily perform the integration and 
obtain the solution for $\cal P$ on the Cauchy horizon as
\begin{equation}
{\cal P}=P_{\rm c}r^{-\beta_{\rm c}}, \label{eq:Cauchy-P-sol}
\end{equation}
where $P_{\rm c}$ is an integration constant and 
$\beta_{\rm c}:=\beta(W_{\rm c})$.
Since $\beta_{\rm c}<0$, $\cal P$ vanishes 
in the limit $r\rightarrow0$.

\subsubsection{Non-Radial Singular Null Geodesics of  
${\cal P}\rightarrow\infty$ for $r\rightarrow0$}

Let us consider non-radial singular null geodesics of 
${\cal P}\rightarrow +\infty$ for $r\rightarrow0$. 
The argument from Eqs.(\ref{eq:W0-eq-1}) 
to (\ref{eq:W0-inequality}) is almost completely applicable to this case. 
However, because $\cal P$ has to diverge for $r\rightarrow0$,  
$\beta$ should be positive. Hence the smaller root 
$W_{0}=W_{-}$ of Eq.(\ref{eq:W0-eq-2}) must be chosen. 

The trajectory in the $(r,\varphi)$-plane is determined by 
the equation
\begin{equation}
{d\varphi\over dr}={l\over R^{2}}
{d\lambda\over dr}
={l(W^{3}+2)r^{\beta_{-}-1}\over 3W^{3}
\sqrt{P^{2}-l^{2}r^{2\beta_{-}}}},
\end{equation}
where $\beta_{-}:=\beta(W_{-})$.
The above equation becomes for $r\rightarrow0$ as 
\begin{equation}
{d\varphi\over dr}\longrightarrow 
{l(W_{-}{}^{3}+2)r^{\beta_{-}-1}\over 3W_{-}{}^{3}P_{0}}.
\end{equation}
Integrating the above equation, we find 
\begin{equation}
\varphi\longrightarrow
{l(W_{-}{}^{3}+2)r^{\beta_{-}}\over 3\beta_{-}W_{-}{}^{3}P_{0}}
+{\rm const.}
\end{equation}
Hence $\varphi$ has a finite limit for $r\rightarrow0$. 
As discussed by Nolan and Mena\cite{ref:NM02}, this result means 
that the singularity from which the non-radial null geodesics 
of $W_{0}=W_{-}$ emanate is foliated by 2-sphere in accordance 
with Christodoulou's argument\cite{ref:christodoulou-2}. 

\subsubsection{Non-Radial Singular Null Geodesics 
with Finite ${\cal P}|_{r=0}$}

Here, we investigate non-radial singular null geodesics 
with a finite limit of $\cal P$ for $r\rightarrow0$. We denote 
this limit by ${\cal P}_{0}$ which should be 
larger than or equal to $l$. However, we can easily see that 
there is no consistent solution of the behavior, 
${\cal P}\rightarrow l$ for $r\rightarrow0$ (see Appendix A). 
Hence hereafter, we assume ${\cal P}_{0}>l$. 

By the assumption, ${\cal P}$ is written in the form 
\begin{equation}
{\cal P}={\cal P}_{0}+\delta{\cal P}(r), \label{eq:P-splitted2}
\end{equation}
where $\delta{\cal P}$ satisfies
\begin{equation}
\lim_{r\rightarrow0}\delta{\cal P}=0. \label{eq:DeltaP-limit2}
\end{equation}
Using Eqs.(\ref{eq:R-asymptotic}), (\ref{eq:P-splitted2}) and 
(\ref{eq:DeltaP-limit2}), Eq.(\ref{eq:R-equation}) becomes 
\begin{equation}
W_{0}{}^{2}=\lim_{r\rightarrow0}{1\over3\alpha W_{0}{}^{2}}
\left(W_{0}-{\Lambda_{0}^{1/2}{\cal P}_{0}\over\sqrt{{\cal P}_{0}^{2}-l^{2}}}
~r^{{1\over2}(1-\alpha)}\right)
\left(W_{0}{}^{3}+2r^{{3\over2}(1-\alpha)}\right).
\end{equation}
In order that the above equation has a positive root $W_{0}$, 
$\alpha$ should be $1/3$ or unity. By the same reason as 
in the previous cases, we find that 
$\alpha=1$ corresponds to the singular null
geodesics, and hence we focus on it. 
Then the above equation leads to 
\begin{equation}
\Lambda_{0}^{1/2}={2W_{0}\left(1-W_{0}^{3}\right)
\sqrt{{\cal P}_{0}^{2}-l^{2}}
\over {\cal P}_{0}\left(W_{0}^{3}+2\right)}.
\label{eq:Lambda1}
\end{equation}
Since $\Lambda_{0}^{1/2}$ is positive, $W_{0}$ is less than unity. 

From Eq.(\ref{eq:P-equation}), we find
\begin{equation}
r{d{\cal P}\over dr}\longrightarrow
-{\Lambda_{0}^{1/2}\over W_{0}{}^{2}}\sqrt{{\cal P}_{0}^{2}-l^{2}}
+{{\cal P}_{0}\over3W_{0}{}^{3}}\left(W_{0}{}^{3}+2\right)~~~~~
{\rm for}~~r\longrightarrow0.
\end{equation}
By the assumption of ${\cal P}\rightarrow{\cal P}_{0}$, the right hand 
side of the above equation should vanish and hence following 
equation is obtained, 
\begin{equation}
\Lambda_{0}^{1/2}={{\cal P}_{0}W_{0}
\left(W_{0}^{3}+2\right)\over 3\sqrt{{\cal P}_{0}^{2}-l^{2}}}.
\label{eq:Lambda2}
\end{equation}
From Eqs.(\ref{eq:Lambda1}) and (\ref{eq:Lambda2}), we obtain
\begin{equation}
{\cal P}_{0}^{2}={6l^{2}(1-W_{0}{}^{3})\over 
2-10W_{0}{}^{3}-W_{0}{}^{6}}, \label{eq:P0-solution}
\end{equation}
Since the right hand side of the 
above equation has to be positive, we obtain the same inequality 
as Eq.(\ref{eq:W0-inequality}). 
It is worthy to notice that if Eq.(\ref{eq:W0-inequality}) holds, 
${\cal P}_{0}^{2}$ is larger than $l^{2}$. Multiplying each side 
of Eqs.(\ref{eq:Lambda1}) and (\ref{eq:Lambda2}), we get
\begin{equation}
\Lambda_{0}=I^{2}(W_{0}):={2\over3}~W_{0}^{2}\left(1-W_{0}{}^{3}\right).
\label{eq:W0-eq-3}
\end{equation}

We also depict the function $I(y)$ in Fig.1. 
From this figure, we can easily see that 
there are two positive real roots of Eq.(\ref{eq:W0-eq-3}). 
Because of Eq.(\ref{eq:W0-inequality}), we have to 
choose the smaller root which will be denoted by $W_{+}$. 
Fig.\ref{fg:H-I} shows that $W_{+}$ is always larger than $W_{-}$. 
Hence the singular null geodesics of $W_{0}=W_{+}$ are  
in the past of the singular null geodesics of $W_{0}=W_{-}$ 
in the sufficiently small neighborhood 
of the central naked singularity $r=0$. 

By the assumption, we may rewrite $W$ in the form
\begin{equation}
W=W_{+}+\delta W(r),
\end{equation}
where 
\begin{equation}
\lim_{r\rightarrow0}\delta W=0.
\end{equation}
Then in the limit of $r\rightarrow0$, 
Eqs.(\ref{eq:P-equation}) and (\ref{eq:R-equation}) are approximated 
by the linearized equations with respect to $(\delta W,~\delta{\cal P})$, 
\begin{equation}
{d\over d\ln r}\left(
\begin{array}{c}
\delta W\\
\delta{\cal P}\\
\end{array}
\right)
=\left(
\begin{array}{cc}
M_{11} & M_{12} \\
M_{21} & M_{22} \\
\end{array}
\right)
\left(
\begin{array}{c}
\delta W\\
\delta{\cal P}\\
\end{array}
\right)
~~, \label{eq:matrix-eq}
\end{equation}
where
\begin{eqnarray}
M_{11}&=&{2-10W_{+}{}^{3}-W_{+}{}^{6}
\over 3W_{+}{}^{3}(W_{+}{}^{3}+2)}, \\
M_{12}&=&
{\sqrt{2(1-W_{+}{}^{3})(2-10W_{+}{}^{3}-W_{+}{}^{6})^{3}}
\over 6\sqrt{3}~l~W_{+}^{2}
(W_{+}{}^{3}+2)^{2}}, \\
M_{21}&=&
{2l~(2W_{+}{}^{3}+1)\sqrt{2(1-W_{+}{}^{3})}\over W_{+}^{4}
\sqrt{3(2-10W_{+}{}^{3}-W_{+}{}^{6})}},\\
M_{22}&=&-{2-10W_{+}{}^{3}-W_{+}{}^{6}
\over 3W_{+}{}^{3}(W_{+}{}^{3}+2)}.
\end{eqnarray}

Maiking the matrix $M_{ij}$ to be the diagonal form, 
Eq.(\ref{eq:matrix-eq}) is written in the form,
\begin{equation}
{d\over d\ln r}\left(
\begin{array}{c}
S_{1}\\
S_{2}\\
\end{array}
\right)
=\left(
\begin{array}{cc}
\gamma & 0 \\
0    & -\gamma \\
\end{array}
\right)
\left(
\begin{array}{c}
S_{1}\\
S_{2}\\
\end{array}
\right)
~~, \label{eq:matrix-eq2}
\end{equation}
where
\begin{equation}
\left(
\begin{array}{c}
S_{1}\\
S_{2}\\
\end{array}
\right)
={1\over 2\gamma M_{12}}\left(
\begin{array}{cc}
M_{11}+\gamma & M_{12} \\
-M_{11}+\gamma & -M_{12}\\
\end{array}
\right)
\left(
\begin{array}{c}
\delta W\\
\delta {\cal P}\\
\end{array}
\right)
~~, 
\end{equation}
and
\begin{equation}
\gamma:=\sqrt{-{\rm det}M}
={\sqrt{(4-8W_{+}{}^{3}-5W_{+}{}^{6})
(2-10W_{+}{}^{3}-W_{+}{}^{6})}
\over 3W_{+}{}^{3}(W_{+}{}^{3}+2)}.
\end{equation}
Since Eq.(\ref{eq:W0-inequality}) should hold, $\gamma$ is real. 

Solving Eq.(\ref{eq:matrix-eq2}), we obtain
\begin{equation}
\left(
\begin{array}{c}
S_{1}\\
S_{2}\\
\end{array}
\right)
=\left(
\begin{array}{c}
C_{1}r^{+\gamma}\\
C_{2}r^{-\gamma}\\
\end{array}
\right)
~~, 
\end{equation}
where $C_{1}$ and $C_{2}$ are integration constants. 
Since $\delta W$ and $\delta{\cal P}$ vanish for $r\rightarrow0$, 
$C_{2}$ should be equal to zero. Hence we find that for $r\rightarrow0$, 
\begin{eqnarray}
W&\longrightarrow& W_{+}+W_{\dagger0}r^{\gamma}, \label{eq:W-asymptotic-2}\\
{\cal P}&\longrightarrow&{\cal P}_{0}-{M_{11}-\gamma \over M_{12}}
~W_{\dagger0}r^{\gamma},\label{eq:P-asymptotic-2}
\end{eqnarray}
where $W_{\dagger0}$ is an arbitrary constant. 

The asymptotic behavior of the trajectory in the $(r,\varphi)$-plane 
is determined by
\begin{equation}
{d\varphi\over dr}
={l(W^{3}+2)\over 3rW^{3}\sqrt{{\cal P}^{2}-l^{2}}}
\longrightarrow{l(W_{+}{}^{3}+2)\over 3rW_{+}{}^{3}
\sqrt{{\cal P}_{0}{}^{2}-l^{2}}}
~~~~~~{\rm for}~~r\longrightarrow0.
\end{equation}
From the above equation, we obtain
\begin{equation}
\varphi
\longrightarrow{l(W_{+}{}^{3}+2)\over 3W_{+}{}^{3}
\sqrt{{\cal P}_{0}{}^{2}-l^{2}}}~\ln r
+{\rm const.}~~~~{\rm for}~~r\longrightarrow0.
\end{equation}
Hence we find that $\varphi\rightarrow-\infty$ in the limit 
$r\rightarrow0$. As discussed by Nolan and Mena\cite{ref:NM02}, 
this means that the singularity 
from which the singular null geodesics of $W_{0}=W_{+}$ emanate 
is topologically a ``point'' in the sense of 
Christodoulou\cite{ref:christodoulou-2}. 
As will be shown below, the singular null geodesics of 
$W_{0}=W_{+}$ have the largest impact parameter at each moment.

\subsection{The Basic Equations for Numerical Calculations}

In order to see the intermediate behavior of singular null geodesics 
inside the dust sphere, we numerically 
solve Eqs.(\ref{eq:P-equation}) and (\ref{eq:R-equation}) with
$\alpha=1$ by the 4th order Runge-Kutta method. 
By virtue of the existence of the homothetic Killing vector, 
we obtain an analytic expression for $\cal P$ as a function of $W$ 
(see Appendix B). However, we numerically integrate 
Eq.(\ref{eq:P-equation}) and 
use the analytic expression for $\cal P$ 
to check the numerical accuracy. 
We will rewrite Eqs.(\ref{eq:P-equation}) and (\ref{eq:R-equation}) 
into appropriate forms for the numerical integrations below. 

\subsubsection{Singular Null Geodesics of 
${\cal P}\rightarrow \infty$ for $r\rightarrow0$}

Here we consider singular null geodesics of 
${\cal P}\rightarrow\infty$ for $r\rightarrow0$, or equivalently 
$W_{0}=W_{-}$. 
From the asymptotic solutions (\ref{eq:W-asymptotic-1}) and 
(\ref{eq:P-asymptotic-1}), we introduce following variables, 
\begin{eqnarray}
w&:=&r^{\beta_{-}}, \\
W_{*}(w)&:=&w^{-1}\Delta W.
\end{eqnarray}
Then Eqs.(\ref{eq:P-equation}) and (\ref{eq:R-equation}) are 
written in the form,
\begin{eqnarray}
{dP\over dw}&=&
{2W_{*}+\left(1+3\beta_{-}\right)F_{1}+3\Lambda_{0}^{1/2}wG_{1}
\over3\beta_{-} (W_{0}+wW_{*})^{4}}~P , \label{eq:P-eq}\\
{dW_{*}\over dw}&=&-{2\left(1+3\beta_{-}\right)W_{*}F_{2}
+\left(2W_{0}+\Lambda_{0}^{1/2}\right)
F_{3}+\Lambda_{0}^{1/2}(W_{0}^{3}+2+wF_{2})G_{2}
\over6\beta_{-} (W_{0}+wW_{*})^{3}}, \label{eq:R-eq}
\end{eqnarray}
where
\begin{eqnarray}
F_{1}&:=&4W_{0}^{3}W_{*}+6wW_{0}^{2}W_{*}^{2}+4w^{2}W_{0}W_{*}^{3}
      +w^{3}W_{*}^{4}, \\
F_{2}&:=&3W_{0}^{2}W_{*}+3wW_{0}W_{*}^{2}+w^{2}W_{*}^{3}, \\
F_{3}&:=&3W_{0}W_{*}^{2}+wW_{*}^{3}, \\
G_{1}&:=&{l^{2}\over P\left(P+\sqrt{P^{2}-l^{2}w^{2}}\right)}, \\
G_{2}&:=&{l^{2}\over \left(P+\sqrt{P^{2}-l^{2}w^{2}}\right)
\sqrt{P^{2}-l^{2}w^{2}}}.
\end{eqnarray}
Eqs.(\ref{eq:P-eq}) and (\ref{eq:R-eq}) can be easily integrated 
numerically from the origin $w=0$, or conversely 
from the surface $w=w_{\rm s}:=r_{\rm s}{}^{\beta_{-}}$ to $w=0$, since 
$w=0$ is not singular in these equations. 

It should be noted that $W$ and $P$ is not uniquely determined
by its initial value $W_{0}$ and $P_{0}$ only. 
Eq.(\ref{eq:R-eq}) shows that 
the boundary value of $W_{*}$ is also necessary to determine them;  
we have denoted the value of $W_{*}(0)$ by $W_{*0}$. 
Since $W_{0}$ should be equal to $W_{-}$, the free parameters 
to specify the solutions for $W$ and $P$ are $W_{*0}$, $P_{0}$ 
and the conserved angular momentum $l$. 
However, here note that the solutions for $W$ constitute 
a two-parameter family with respect to $W_{*0}$ and $l/P_{0}$.  
We can rewrite Eqs.(\ref{eq:P-eq}) and (\ref{eq:R-eq}) as the 
equations for ${\hat P}:=P/P_{0}$ and $W_{*}$. 
By introducing ${\hat l}:=l/P_{0}$, $P_{0}$ does not appear in 
these rewritten equations. The initial value of ${\hat P}$ 
is unity by definition. Hence the free parameters are 
$W_{*0}$ and ${\hat l}$ in this new system of differential equations.

\subsubsection{Non-Radial Singular Null Geodesics 
with Finite ${\cal P}|_{r=0}$}

We consider the case of ${\cal P}\rightarrow$finite for 
$r\rightarrow0$, or equivalently $W_{0}=W_{+}$. 
In this case, the conserved angular momentum $l$ is necessarily 
non-vanishing.  From Eqs.(\ref{eq:W-asymptotic-2}) and 
(\ref{eq:P-asymptotic-2}), it is better to introduce following variables,
\begin{eqnarray}
v&:=&r^{\gamma}, \\
W_{\dagger}(v)&:=&v^{-1}\delta W, \\
{\cal P}_{\dagger}(v)&:=&v^{-1}\delta{\cal P}.
\end{eqnarray}
Then Eqs.(\ref{eq:P-equation}) and (\ref{eq:R-equation}) are written 
in the form,
\begin{eqnarray}
{d{\cal P}_{\dagger}\over dv}&=&
{1\over 3\gamma W^{4}v^{2}}\left\{W(W^{3}+2)-3\Lambda_{0}^{1/2}
\sqrt{{\cal P}^{2}-l^{2}}-3\gamma W^{4}{\cal P}_{\dagger}v\right\}, 
\label{eq:Pdagger-eq}\\
{dW_{\dagger}\over dv}&=&{1\over 6\gamma W^{3}v^{2}}
\left\{2W(1-W^{3})-{\Lambda_{0}^{1/2}{\cal P}(W^{3}+2)\over 
\sqrt{{\cal P}^{2}-l^{2}}}-6\gamma W^{3}W_{\dagger}v\right\}.
\label{eq:Wdagger-eq}
\end{eqnarray}
Eqs.(\ref{eq:W-asymptotic-2}) and (\ref{eq:P-asymptotic-2}) impose 
boundary conditions on $W_{\dagger}$ and ${\cal P}_{\dagger}$ as 
\begin{eqnarray}
W_{\dagger}(0)&=& W_{\dagger 0}, \label{eq:Wdagger-initial}\\
{\cal P}_{\dagger}(0)&=& {(\gamma-M_{11})W_{\dagger 0}\over M_{12}},
\label{eq:P-initial}
\end{eqnarray}
where $W_{\dagger 0}$ is an arbitrary constant. 
The right hand sides of Eqs.(\ref{eq:Pdagger-eq}) and 
(\ref{eq:Wdagger-eq}) seem to be singular at 
$v=0$, but it is not true. Taking into account 
Eqs.(\ref{eq:Wdagger-initial}) and (\ref{eq:P-initial}), we 
can verify that the right hand sides 
have finite limits for $v\rightarrow0$. 

Once the initial value $W_{\dagger 0}$ is given, 
the initial value of ${\cal P}_{\dagger}$ is uniquely fixed 
by Eq.(\ref{eq:P-initial}). Hence the solutions of $W_{\dagger}(v)$ 
constitute an one-parameter family with respect to $W_{\dagger 0}$ 
(the solutions of $W_{\dagger}$ seem to constitute a 
two-parameter family with respect to $W_{\dagger0}$ 
and $l$, but it is not true). 
In order to see this fact, let us introduce following variables
\begin{eqnarray}
{\tilde {\cal P}}_{0}&:=&{{\cal P}_{0}\over l}
=\sqrt{6(1-W_{+}{}^{3})\over 
2-10W_{+}{}^{3}-W_{+}{}^{6}}, \label{eq:p0-initial}\\
{\tilde{\cal P}}_{\dagger}&:=&{{\cal P}_{\dagger}\over l},
\end{eqnarray}
and rewrite Eqs.(\ref{eq:Pdagger-eq}) and (\ref{eq:Wdagger-eq}) 
into the equations for these variables.  
Then the parameter $l$ disappears in the new equations. 
This means that the solutions for ${\tilde {\cal P}}_{\dagger}$ and 
$W_{\dagger}$ do not depend on $l$. 
Since ${\tilde {\cal P}}_{0}$ is not an arbitrary 
constant by Eq.(\ref{eq:p0-initial}), 
the remaining free parameter is the initial value $W_{\dagger0}$ only.

\section{Null Geodesics in the Exterior Region}

The components of the null geodesic tangent on the equatorial 
plane $\theta=\pi/2$  in the exterior Schwarzschild static chart 
are given by
\begin{eqnarray}
k^{T}&=&{dT\over d\lambda}={\omega\over C(R)}, \\
k^{R}&=&{dR\over d\lambda}=
\omega\sqrt{1-{C(R)\over R^{2}}\left({l\over \omega}\right)^{2}}, 
\label{eq:k-R-component}\\
k^{\theta}&=&{d\theta\over d\lambda}=0, \\
k^{\varphi}&=&{d\varphi\over d\lambda}={l\over R^{2}},
\end{eqnarray}
where $\omega$ and $l$ are integration constants which correspond 
to the angular frequency measured by the static observer at the future 
null infinity and to the conserved angular momentum, respectively. 

In order to obtain a relation between the components of the interior 
and of the exterior regions, we introduce a 
tetrad basis whose components with respect to the interior 
LTB solution are given by 
\begin{eqnarray}
\left(e_{(t)}{}^{\mu}\right)_{\rm LTB}&=&\left(1,~0,~0,~0\right), \\
\left(e_{(r)}{}^{\mu}\right)_{\rm LTB}&=&\left(0,~{1\over R'},~0,~0\right),
\label{eq:r-component}\\
\left(e_{(\theta)}{}^{\mu}\right)_{\rm LTB}&=&\left(0,~0,~{1\over R},~0
\right),\\
\left(e_{(\varphi)}{}^{\mu}\right)_{\rm LTB}&=&\left(0,~0,~0,~
{1\over R\sin\theta}\right).\label{eq:phi-component}
\end{eqnarray}
Note that $e_{(t)}{}^{\mu}$ agrees with the 4-velocity of the 
dust fluid element and $e_{(r)}{}^{\mu}$ is normal to 
$e_{(t)}{}^{\mu}$ in $(t,r)$-space. 
At the surface of the dust sphere $r=r_{\rm s}$, the components 
of $e_{(t)}{}^{\mu}$ with respect to 
the exterior Schwarzschild static chart are obtained by 
Eqs.(\ref{eq:u0-sol}) and (\ref{eq:u1-sol}). The components 
of $e_{(r)}{}^{\mu}$ with respect to the exterior static chart 
are obtained as an outward normal to $e_{(t)}{}^{\mu}$ in $(T,R)$-space. 
Hence the components of the tetrad basis with respect to the 
exterior static chart are given by
\begin{eqnarray}
\left(e_{(t)}{}^{\mu}\right)_{\rm Sch}&=&\left({1\over C(R)},
~-\sqrt{2M\over R},
~0,~0\right), \\
\left(e_{(r)}{}^{\mu}\right)_{\rm Sch}&=&\left(\sqrt{2M\over R},
~{1\over C(R)},~0,~0\right),\\
\left(e_{(\theta)}{}^{\mu}\right)_{\rm Sch}&=&\left(0,~0,~{1\over R},~0
\right),\\
\left(e_{(\varphi)}{}^{\mu}\right)_{\rm Sch}&=&\left(0,~0,~0,~
{1\over R\sin\theta}\right). 
\end{eqnarray}
Using the above expression, 
the continuities of $e_{(t)}{}^{\mu}k_{\mu}$ and 
$e_{(r)}{}^{\mu}k_{\mu}$ lead to
the relation between the components in the exterior and in the interior 
coordinate systems at the surface of the dust sphere;
\begin{eqnarray}
k^{T}&=&{1\over C(R)}\left(k^{t}-R'\sqrt{2M\over R}~k^{r}\right), 
\label{eq:kt-junction}\\
k^{R}&=&-\sqrt{2M\over R}~k^{t}+R'k^{r}. \label{eq:kr-junction}
\end{eqnarray}
Therefore, we obtain the angular frequency for the observer 
at the null infinity as 
\begin{equation}
\omega=\left(k^{t}-R'\sqrt{2M\over R}~k^{r}\right)_{r=r_{\rm s}}. 
\label{eq:omega-eq}
\end{equation}
The $\varphi$-components $k^{\varphi}$ in exterior and interior 
coordinate systems agree with each other and hence $l$ is common. 

We consider the situation in which a distant static observer at 
some given areal radius $R=R_{\rm o}$ detects the null geodesics. 
There is the invariance of the time translation, $T\rightarrow T+$const., 
in the exterior region. 
Using this symmetry, we set the origin of the time coordinate $T$ so that 
the observer intersects the Cauchy horizon at $T=0$. 
The future directed outgoing radial null geodesic 
in the exterior Schwarzschild region is determined by
\begin{equation}
{dT\over dR}={1\over C(R)}
\end{equation}
We can easily integrate the above equation and obtain
\begin{equation}
T-T_{\rm ng}=R+2M\ln|R-2M|,
\end{equation}
where $T_{\rm ng}$ is an integration constant. 
Since the Cauchy horizon is generated by radial 
null geodesics which go through 
$R=R_{\rm o}$ at $T=0$, the trajectory of the Cauchy horizon should be 
written in the form
\begin{equation}
T=T_{\rm CH}(R):=R-R_{\rm o}
+2M\ln\Biggl|{R-2M\over R_{\rm o}-2M}\Biggr|. \label{eq:CH-exterior}
\end{equation}
The Cauchy horizon intersects the surface of the dust sphere 
when its areal radius is given by  
\begin{equation}
R=R_{\rm cs}:=r_{\rm s}W_{\rm c}{}^{2}={2MW_{\rm c}{}^{2}\over\Lambda_{0}}. 
\end{equation}
Hence the exterior time 
coordinate at the intersection $R=R_{\rm cs}$ is given by 
\begin{equation}
T=T_{\rm cs}:=T_{\rm CH}(R_{\rm cs}). 
\end{equation}

The trajectory of the surface of the dust sphere 
is given by Eq.(\ref{eq:DS-trajectory}). 
The integration constant $T_{\rm d}$ in this equation 
is determined by the condition 
in which the areal radius of the dust surface is equal to $R_{\rm cs}$ 
at $T=T_{\rm cs}$. Then we obtain
\begin{equation}
T_{\rm d}=T_{\rm cs}+{2\over 3}\sqrt{R_{\rm cs}\over 2M}
\left(R_{\rm cs}+6M\right)
+2M\ln\Biggl|{\sqrt{R_{\rm cs}}-\sqrt{2M}
\over \sqrt{R_{\rm cs}}+\sqrt{2M}}\Biggr|,
\end{equation}

To obtain non-radial singular null geodesics 
in the exterior region, we integrate the null condition  
\begin{equation}
{dT\over dR}={R\over C(R)\sqrt{R^{2}-(l/\omega)^{2}C(R)}}, 
\label{eq:null-condition}
\end{equation}
between the surface of the dust sphere $R=R_{\rm s}$ and the observer 
$R=R_{\rm o}$.

\section{Apparent Size of The Naked Singularity}

First we consider the angle $\delta$ between 
the radial direction and the singular null geodesic at the observer 
$R=R_{\rm o}$ (see Fig.\ref{fg:angle}). This angle $\delta$ 
is defined as follows. We introduce an 
orthonormal basis whose components with respect to the 
Schwarzschild static chart are given by 
\begin{eqnarray}
(e_{(T)}{}^{\mu})_{\rm Sch}&=&\left({1\over\sqrt{C(R_{\rm o})}},
~0,~0,~0\right), \\
(e_{(R)}{}^{\mu})_{\rm Sch}&=&\left(0,~\sqrt{C(R_{\rm o})},~0,~0\right), \\
(e_{(\theta)}{}^{\mu})_{\rm Sch}&=&\left(0,~0,~{1\over R_{\rm o}},~0\right), \\
(e_{(\varphi)}{}^{\mu})_{\rm Sch}
&=&\left(0,~0,~0,{1\over R_{\rm o}\sin\theta}\right).
\end{eqnarray}
Then the angle $\delta$ is given by
\begin{equation}
\tan{\delta}
={e_{(\varphi)}{}^{\mu}k_{\mu}\over e_{(R)}{}^{\mu}k_{\mu} }
={l\over \omega}\sqrt{C(R_{\rm o})\over R_{\rm o}{}^{2}
-(l/\omega)^{2}C(R_{\rm o})}.
\end{equation}
We can easily see that for sufficiently large $R_{\rm o}$, 
the angle $\delta$ becomes as
\begin{equation}
\delta\longrightarrow {b\over R_{\rm o}},
\end{equation}
where $b$ is a constant which is regarded as 
the impact parameter of the singular null geodesic 
with respect to the central naked singularity, 
\begin{equation}
b:=\lim_{R_{\rm o}\rightarrow\infty} R_{\rm o}\delta
={l\over \omega}, \label{eq:b-def}
\end{equation}
The angle $\delta$ depends on the location $R_{\rm o}$ of the observer. 
On the other hand, the impact parameter $b$ does not and hence 
it is more convenient to characterize the ``size'' of the naked 
singularity. 

Eq.(\ref{eq:k-R-component}) is regarded as an energy equation 
of a test particle with a specific energy $1/2$,
\begin{equation}
{1\over2}\left({dR\over d{\tilde \lambda}}\right)^{2}+V(R;b)={1\over2}.
\end{equation}
where ${\tilde \lambda}\equiv \omega\lambda$ and
\begin{equation}
V(R;b)\equiv {b^{2}C(R)\over 2R^{2}}.
\end{equation}
The maximal value of the effective potential $V$ is 
$b^{2}/54M^{2}$ at $R=3M$. 
Hence if the impact parameter $b$ is smaller than $3\sqrt{3}M$, 
there is no forbidden region for 
the motion of the test particle 
and hence the null geodesic can escape to infinity if 
it is initially outgoing. 
On the other hand, in the case of $b\geq 3\sqrt{3}M$, 
it is non-trivial whether the singular null geodesics 
can go away to the infinity. 
When $b$ is equal to $3\sqrt{3}M$, the null geodesic 
of the circular orbit $R=3M$ is possible although 
such an orbit is unstable. 

\begin{figure}
        \centerline{\epsfxsize 10cm \epsfysize 7cm \epsfbox{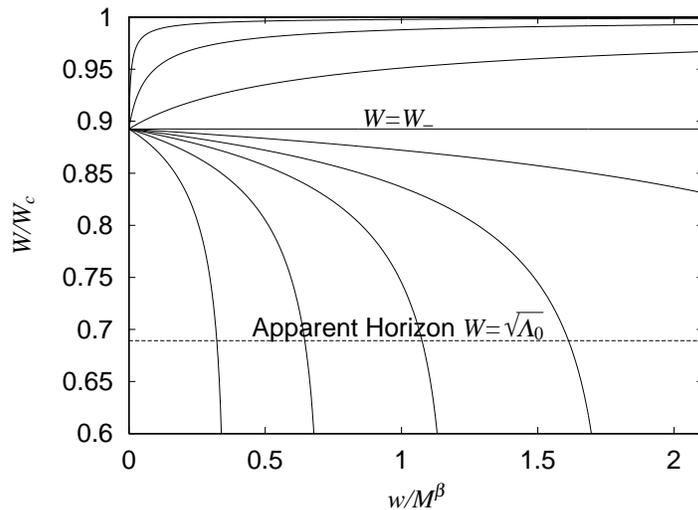}}
        \caption{The trajectories of radial singular null geodesics 
$l=0$ are depicted in $(w,W)$-plane. 
From this figure, we can see that in the limit of 
$W_{*0}\rightarrow\infty$,  
the singular null geodesic approaches to 
the Cauchy horizon $W=W_{\rm c}$. We also plot the apparent horizon 
$F/R=1$, or equivalently $W=\sqrt{\Lambda_{0}}$ in this plane. 
} 
\label{fg:radial}
\end{figure}

\begin{figure}
        \centerline{\epsfxsize 10cm \epsfysize 7cm \epsfbox{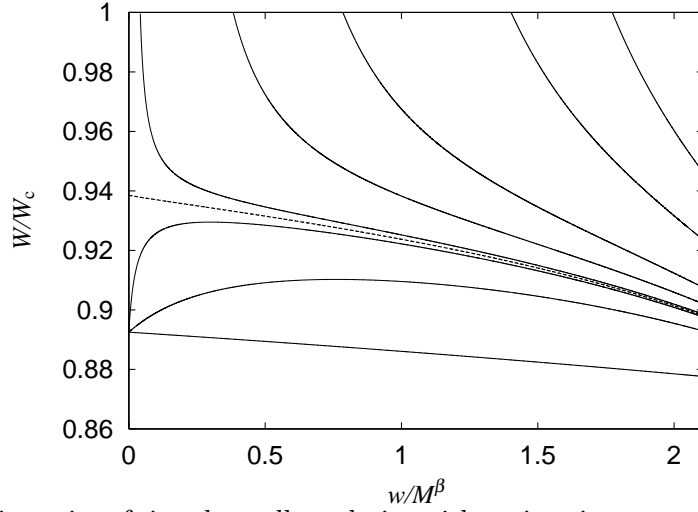}}
        \caption{The trajectories of singular null geodesics 
with various impact parameters $b$ are depicted in $(w,W)$-plane. 
These null geodesics intersect with each other at $(T,R)=(5M,100M)$ 
in the exterior region.  
The lower most curve corresponds to the radial null geodesic $l=0$ ($b=0$). 
The dashed curve is the non-radial singular null geodesic with the 
largest impact parameter $b=b_{\rm crit}$. 
The solid curves below this dashed curve are all 
the singular null geodesics while those above this curve 
are not. 
} 
\label{fg:non-radial}
\end{figure}

\begin{figure}
        \centerline{\epsfxsize 13cm \epsfysize 5cm \epsfbox{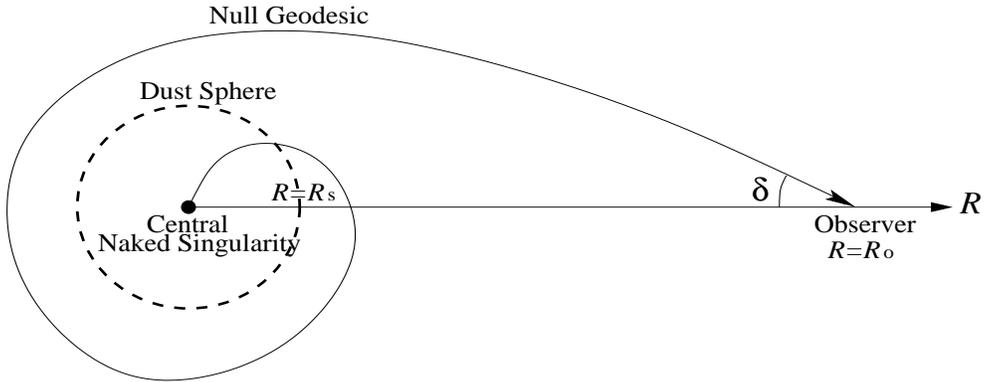}}
        \caption{The schematic diagram of the trajectory of a 
 non-radial singular null geodesic.  
} 
\label{fg:angle}
\end{figure}

\begin{figure}
        \centerline{\epsfxsize 13cm \epsfysize 9.05cm \epsfbox{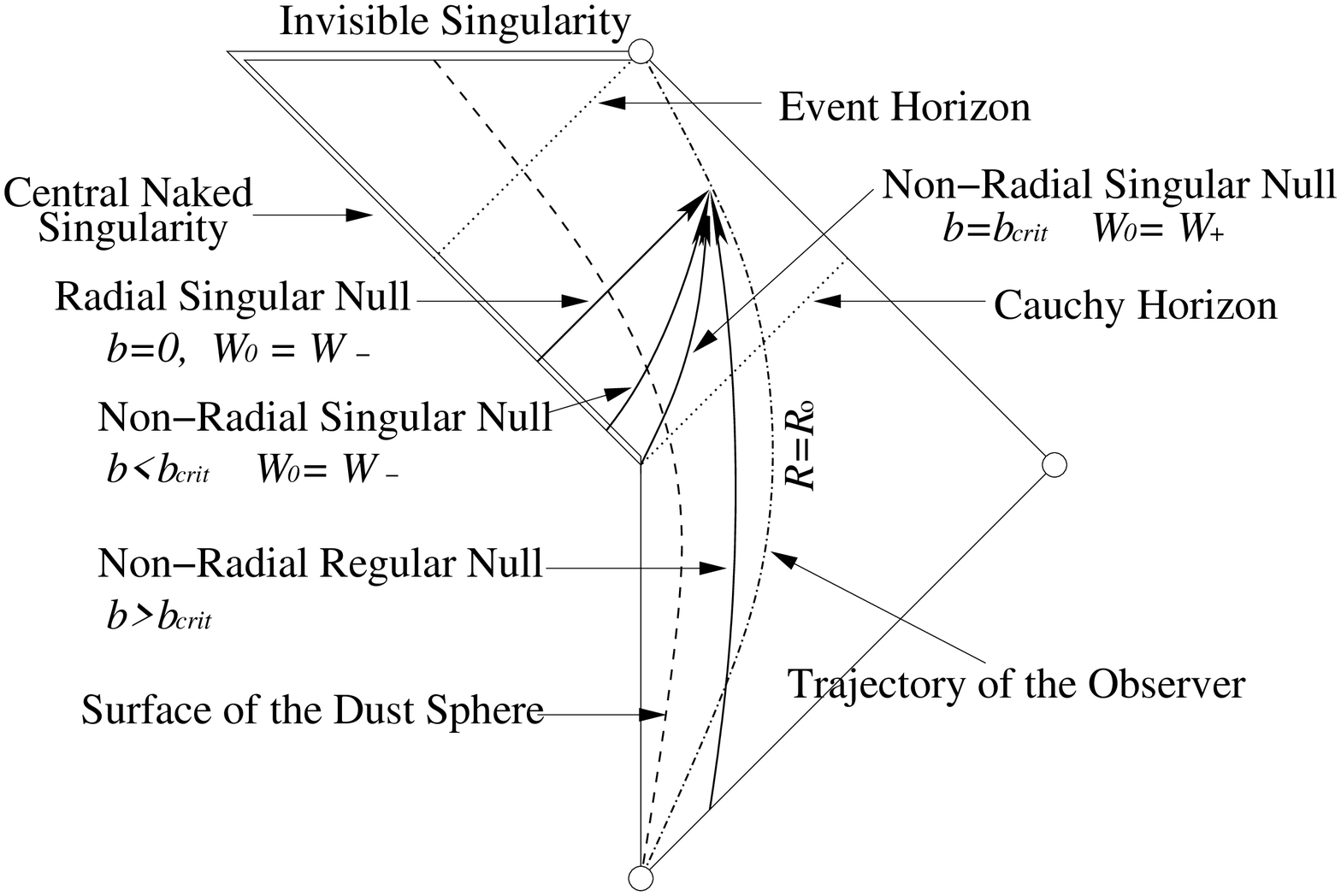}}
        \caption{We depict the conformal diagram of the spacetime with 
a self-similar dust sphere. The singular null geodesic of 
$b=b_{\rm crit}$ emanates from the beginning of Cauchy horizon 
in this diagram. The observed null geodesics of $b>b_{\rm crit}$ 
do not go through the origin $r=0$. 
} 
\label{fg:causal}
\end{figure}

Numerically integrating geodesic equations derived in Sec.III, 
we can draw trajectories in the $(w,W)$-plane. 
As an example, we depict the radial singular null geodesics  
of $W_{-}=\sqrt{3/10}$ in Fig.\ref{fg:radial}. 
By Eq.(\ref{eq:W0-eq-2}), $\Lambda_{0}$ is about $0.288$. 
The radial singular null geodesic of $W_{0}=W_{-}$ 
approaches to the Cauchy horizon $W=W_{\rm c}$ in the limit of  
$W_{*0}\rightarrow\infty$.  
On the other hand, in Fig.\ref{fg:non-radial}, we depict the trajectories 
of the non-radial singular null geodesics of $W_{0}=W_{\pm}$. 
As in Fig.\ref{fg:radial}, we choose $W_{-}=\sqrt{3/10}$. 
In this case, $W_{+}$ is about $0.361$. 
All the null geodesics in this figure are detected 
by the observer at $R_{\rm o}=100M$ at the moment $T= 5M$. 
Hence these null geodesics constitute an one-parameter family 
with respect to the impact parameter $b$. 
The lower most curve corresponds to the radial singular null geodesic 
$b=0$. There is a critical value 
$b_{\rm crit}$ for the impact parameter $b$  
($b_{\rm crit}$ is about $0.460\times 3\sqrt{3}$ in the case of this figure). 
The null geodesics of $b>b_{\rm crit}$ do not start from $w=0$ 
and hence are not singular null geodesics; $W$ of these null geodesics 
becomes larger than $W_{\rm c}$ near $w=0$. 
On the other hand, the null geodesics of $b\leq b_{\rm crit}$ 
start from the singularity $w=0$. For $b<b_{\rm crit}$, 
$W$ is equal to $W_{-}$ at $w=0$,
while in the case of $b=b_{\rm crit}$, $W$ becomes $W_{+}$ 
at $w=0$. Hence the singular null geodesic of $W_{0}=W_{+}$ 
has the largest impact parameter(see also Fig.\ref{fg:causal}). 

The angular diameter $\Delta$ of the central naked singularity 
for an observer at $R=R_{\rm o}$ 
is the twice of the angle $\delta$ of the singular 
null geodesic with the largest impact parameter $b_{\rm crit}$,
\begin{equation}
\Delta=2\arctan \left[b_{\rm crit}\sqrt{C(R_{\rm o})\over R_{\rm o}{}^{2}
-b_{\rm crit}{}^{2}C(R_{\rm o})}\right].
\end{equation}

In order to study the temporal behavior of the largest impact parameter 
$b=b_{\rm crit}$, we integrate Eqs.(\ref{eq:Pdagger-eq}) 
and (\ref{eq:Wdagger-eq}) from $v=0$ to $v=r_{\rm s}{}^{\gamma}$ 
for various boundary values $W_{\dagger0}$. We set $l$ to be unity. 
After the integration to $v=r_{\rm s}{}^{\gamma}$ is completed, 
we obtain $\omega$ by Eq.(\ref{eq:omega-eq}); we denote 
$\omega$ of the null geodesic of $b=b_{\rm crit}$ by 
$\omega_{\rm crit}$. Then we 
integrate Eq.(\ref{eq:null-condition}) until $R=R_{\rm o}$ and  
as a result, we obtain $T_{\rm o}$. 
Here note that $T_{\rm o}$ is a function of the 
initial value $W_{\dagger0}$.  We search for the initial value 
$W_{\dagger0}$ so that some given $T_{\rm o}$ is realized as 
a result of the numerical integration. 
Then we obtain the largest impact parameter 
by $b_{\rm crit}=\omega_{\rm crit}$ for given $T_{\rm o}$. 

\begin{figure}
        \centerline{\epsfxsize 10cm \epsfysize 7cm \epsfbox{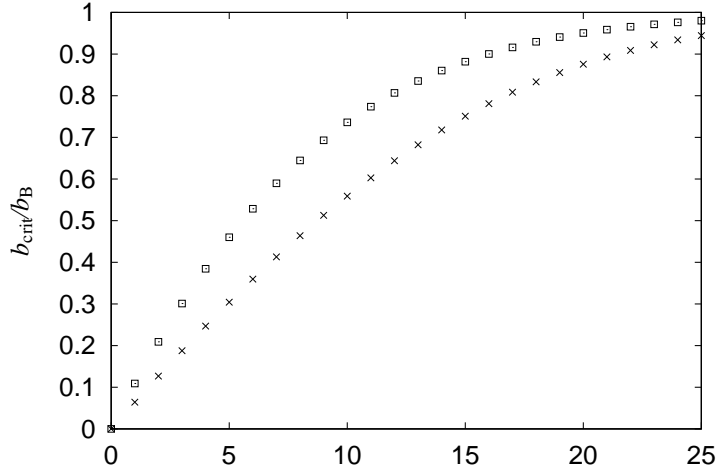}}
        \caption{We depict the impact parameter $b_{\rm crit}$ 
as a function of time $T_{\rm o}$ at the observer $R_{\rm o}=100M$. 
The crosses are the data for the background self-similar 
dust chosen so that $W_{-}=\sqrt{2/10}$ while the squares correspond 
to the data of the background with $W_{-}=\sqrt{3/10}$. 
The impact parameter $b_{\rm crit}$ grows monotonically 
and approaches to the value $b_{\rm B}=3\sqrt{3}M$.
} 
\label{fg:impact}
\end{figure}

In Fig.\ref{fg:impact}, we depict the largest impact parameter 
$b_{\rm crit}$ as a function of the exterior 
time coordinate $T_{\rm o}$ at the observer $R_{\rm o}=100M$. 
Since $R_{\rm o}$ is sufficiently larger 
than $2M$, the angular diameter $\Delta$ of the central naked 
singularity is almost equal to $2b_{\rm crit}/R_{\rm o}$. 

From Fig.\ref{fg:impact}, we see that the largest impact parameter 
$b_{\rm crit}$ approaches to $b_{\rm B}:=3\sqrt{3}M$ asymptotically.   
Hence our numerical results strongly suggest that 
the largest impact parameter $b_{\rm crit}$ is bounded above by 
$b_{\rm B}$. The angular diameter $\Delta$ is also bounded above by 
\begin{equation}
\Delta_{\rm B}=
2\arctan \left[b_{\rm B}\sqrt{C(R_{\rm o})\over R_{\rm o}{}^{2}
-b_{\rm B}{}^{2}C(R_{\rm o})}\right]~. 
\end{equation}
First, at the Cauchy horizon $T_{\rm o}=0$, the central 
naked singularity is observed as a point, i.e., $b_{\rm crit}=0$, or 
equivalently $\Delta=0$. Then gradually the 
largest impact parameter $b_{\rm crit}$ grows and 
approaches to $b_{\rm B}$; correspondingly, the angular diameter 
approaches to $\Delta_{\rm B}$.

\section{Frequencies of Physical Fields from the Naked Singularity}

The redshift $z_{0}$ between the 
central naked singularity and a timelike observer with a 4-velocity 
$v^{\mu}$ is given by
\begin{equation}
z_{0}+1\equiv 
{1\over k_{\mu}v^{\mu}}
\lim_{r\rightarrow 0} k_{\nu}u^{\nu}.
\end{equation}
Since $k_{\nu}u^{\nu}=-{\cal P}/R\rightarrow-\infty$ for 
$r\rightarrow0$ for finite $k_{\mu}v^{\mu}$, 
the redshift $z_{0}$ is infinite. 
From this fact, it seems to be likely that physical 
information of the naked singularity cannot arrive at any observer. 
However it may not be necessarily true. 
Since the energy density and spacetime curvature are infinite 
at the naked singularity, physical fields generated there
might also have infinite frequencies. 
Hence even if the physical fields 
suffer the infinite redshift, it is a non-trivial issue whether 
their frequencies observed at infinity vanish or not. 

We do not exactly know what and 
how amount of physical fields are generated in the neighborhood of 
the central naked singularity where quantum effects of gravity 
might be important. However we may say that 
the physical field which can arrive at the 
distant observer from the dense region around the naked singularity 
might be graviton due to the weakness of its coupling to other 
fields. Further, it might not be so terrible to assume that 
the frequency of the graviton is order of $\sqrt{\rho}$ which is the  
inverse of the free fall time. 
From Eq.(\ref{eq:rho-sol}), along a singular null geodesic, 
the density $\rho$ behaves as
\begin{equation}
\rho\longrightarrow 
{3W_{0}{}^{3}(1-W_{0}{}^{3})^{2}\over 2\pi(W_{0}{}^{3}+2)^{3}}
\times {1\over R^{2}}~~~~~{\rm for}~~r\longrightarrow0.
\end{equation}
Hence hereafter we will focus on a physical field which 
is generated at $R=L$ with a 
comoving angular frequency $\omega_{L}=1/L$. 
Of course, $L$ is assumed to be sufficiently smaller than $M$. 
It should be noted that the angular frequency of the physical 
field propagating along such a null geodesic 
becomes infinite at the central naked singularity. 

\subsection{On the Cauchy Horizon}

First, let us consider the angular frequency of the physical 
field which propagates on the Cauchy horizon. 
As mentioned, the Cauchy horizon is $R=W_{\rm c}{}^{2}r$.   
Hereafter we will denote the comoving radius of 
the singular null geodesic at $R=L$ by $r_{L}$. 
Then $r_{L}$ is equal to $L/W_{\rm c}{}^{2}$ on the Cauchy horizon. 
From Eq.(\ref{eq:Cauchy-P-sol}), the angular frequency at $R=L$ 
is given by  
\begin{equation}
\omega_{L}
={P_{\rm c}W_{\rm c}{}^{2\beta_{\rm c}}\over L^{\beta_{\rm c}+1}}~.
\end{equation}
As mentioned, we assume that the angular frequency $\omega_{L}$  
is $1/L$ at $R=L$ and hence from the above equation, we obtain  
\begin{equation}
P_{\rm c}=\left({L\over W_{\rm c}{}^{2}}\right)^{\beta_{\rm c}}.
\end{equation}
From Eq.(\ref{eq:omega-eq}), the angular frequency  
for a static observer at infinity is given by 
\begin{eqnarray}
\omega
&=&\left[k^{t}-R'\sqrt{2M\over R}~k^{r}\right]_{r=r_{\rm s}} \nonumber \\
&=&{1\over 2M}\left({L\over 2M}\right)^{\beta_{\rm c}}
\left\{{4W_{\rm c}(1-W_{\rm c}{}^{3})^{2}\over (W_{\rm c}+2)^{2}}
\right\}^{\beta_{\rm c}+1}
\left\{1+{2W_{\rm c}{}^{1/2}(1-W_{\rm c}{}^{3})\over W_{\rm c}+2}
\right\}.
\end{eqnarray}
It should be noted that since $\beta_{\rm c}$ is negative, 
$\omega$ diverges in the limit of $L\rightarrow0$. 

We find that in the limit of $W_{\rm c}\rightarrow 1$, 
the angular frequency $\omega$ behaves as
\begin{equation}
\omega \longrightarrow {1\over L}~.
\end{equation}
This is because $W_{\rm c}=1$ corresponds to 
no dust-sphere case $\Lambda_{0}=0$. Hence in the case of 
sufficiently small $\Lambda_{0}$, the angular frequency 
$\omega$ observed at the infinity is 
almost equal to the initial value $\omega_{L}$ at $R=L$. 
However it should be noted that small $\Lambda_{0}$ does not 
necessarily corresponds to small total mass $M$. 
Because of $M=\Lambda_{0}r_{\rm s}$, if the comoving radius 
$r_{\rm s}$ of the dust surface is very large, $M$ can also be large. 
On the other hand, in the limit of $W_{\rm c}\rightarrow W_{\rm m}$, 
the angular frequency becomes as
\begin{equation}
\omega\longrightarrow {\Lambda_{\rm crit}\over 2MW_{\rm m}}
\left\{1+\left({\Lambda_{\rm crit}\over W_{\rm m}}\right)^{1/2}\right\}
\sim 0.243M^{-1}.
\end{equation}
The angular frequency $\omega$ does not depend on $L$ in this limit. 

\subsection{In the Future of the Cauchy Horizon}

Now let us consider singular null geodesics in the future of 
the Cauchy horizon. 
The angular frequency of the physical field propagating along 
the singular null geodesic of the 
largest impact parameter $b=b_{\rm crit}$ is obtained  by integrating 
Eqs.(\ref{eq:Pdagger-eq}) and (\ref{eq:Wdagger-eq}) from $v=0$ 
to $v=r_{\rm s}{}^{\gamma}$. 
For the exterior region, we integrate Eq.(\ref{eq:null-condition}) 
until $R$ becomes some given $R_{\rm o}$ to obtain $T$ at 
$R=R_{\rm o}$. Then the initial value of  
$W_{\dagger}$ is determined iteratively 
so that $T$ becomes some given $T_{\rm o}$ at $R=R_{\rm o}$. 
The initial value of $\cal P$ is 
unity, i.e., ${\cal P}_{0}=1$, in order that $\omega_{L}=1/L$. 
This condition and Eq.(\ref{eq:P0-solution}) lead to 
\begin{equation}
l=l_{\rm crit}:=\sqrt{2-10W_{+}{}^{3}-W_{+}{}^{6}\over 6(1-W_{+}{}^{3})}.
\end{equation}
As a result of the numerical integration, we obtain the angular
frequency $\omega_{\rm crit}$ and the impact parameter 
$b_{\rm crit}=l_{\rm crit}/\omega_{\rm crit}$.
Here note that the angular frequency $\omega_{\rm crit}$  
does not depend on the assumed value of $L$. 

In order to study the angular frequencies for the distant static observer 
in the case of the null geodesics of $W_{0}=W_{-}$, 
we had better integrating the geodesic equations from the observer 
toward the central naked singularity. 
First, we integrate Eq.(\ref{eq:null-condition}) from 
the observer $R=R_{\rm o}$ toward the dust sphere $R=R_{\rm s}$. 
The initial condition is the given $T_{\rm o}$. 
We set $\omega$ to be unity and hence the impact parameter 
is equal to $l$ which should be smaller than $b_{\rm crit}$ 
obtained in the above. 
After the null geodesic reaches $R=R_{\rm s}$, 
we integrate Eqs.(\ref{eq:P-eq}) and (\ref{eq:R-eq}) 
toward the central naked singularity $w=0$. 
The initial condition for these equations are obtained by 
Eqs.(\ref{eq:kt-junction}) and (\ref{eq:kr-junction}) as
\begin{equation}
P(w_{\rm s})={w_{\rm s}R_{\rm s}\over C(R_{\rm s})}
\left[1+\sqrt{{2M\over R}
\left\{1-{l^{2}C(R_{\rm s})\over \omega^{2}R_{\rm s}{}^{2}}\right\}}\right],
\end{equation}
and
\begin{equation}
W_{*}(w_{\rm s})={1\over w_{\rm s}}
\left(\sqrt{\Lambda_{0}R_{\rm s}\over 2M}-W_{-}\right).
\end{equation}
Once we obtain $P$ at $R=L$, we obtain the observed angular frequency 
$\omega_{\rm o}$ for the static observer at $R=R_{\rm o}$ as 
\begin{equation}
\omega_{\rm o}={w_{L}\over P(w_{L})\sqrt{C(R_{\rm o})}}, 
\end{equation}
where $w_{L}=r_{L}{}^{\beta_{-}}$. 

\begin{figure}
        \centerline{\epsfxsize 10cm \epsfysize 7cm \epsfbox{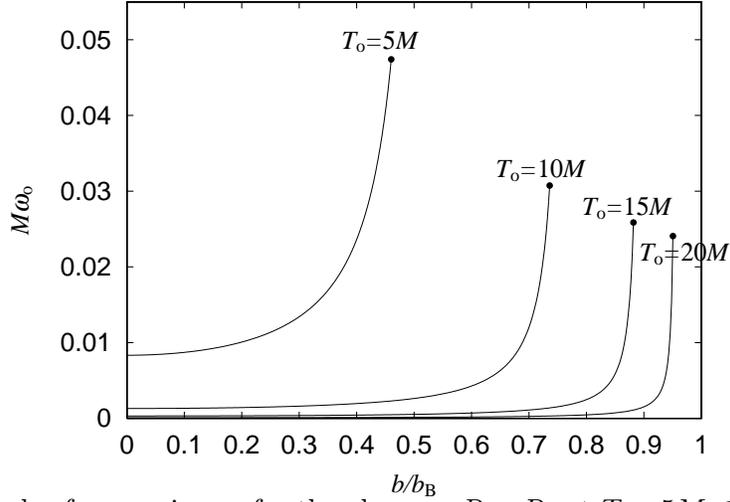}}
        \caption{The angular frequencies $\omega_{\rm o}$ 
for the observer $R=R_{\rm o}$ at $T=5M$, $10M$, $15M$ 
and $20M$ are depicted as a function of the impact parameter $b$. 
The null geodesics are assumed to 
start at $R=10^{-2}M$ with the comoving angular frequency
 $\omega_{L}=10^{2}M^{-1}$. 
The background self-similar dust is chosen so that $W_{-}=\sqrt{3/10}$.
The angular frequency $\omega_{\rm o}$ 
is a monotonically increasing function of $b$. 
} 
\label{fg:omega-1}
\end{figure}

\begin{figure}
        \centerline{\epsfxsize 10cm \epsfysize 7cm \epsfbox{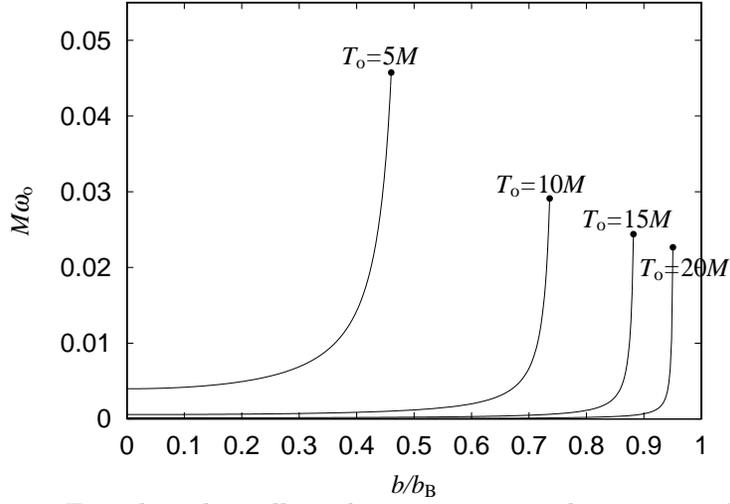}}
        \caption{The same as Fig.\ref{fg:omega-1} but 
the null geodesics are assumed to start at $R=10^{-3}M$ 
with the comoving angular frequency
 $\omega_{L}=10^{3}M^{-1}$. 
} 
\label{fg:omega-2}
\end{figure}

In Figs.\ref{fg:omega-1} and \ref{fg:omega-2}, 
we depict the angular frequency $\omega_{\rm o}$ for the static observer 
at $R_{\rm o}=100M$ as a function of the impact parameter $b$. 
Fig.\ref{fg:omega-1} corresponds to the case of $L=10^{-2}M$ 
while Fig.\ref{fg:omega-2} is for $L=10^{-3}M$. 
Since $R_{\rm o}$ is sufficiently large, 
the angle $\delta$ is almost equal to $b/R_{\rm o}$. 
The background model is chosen so that $W_{-}=\sqrt{3/10}$. 
There are four curves; these are the data at $T_{\rm o}=5M$, 
$10M$, $15M$ and $20M$, respectively. 
The largest impact parameter $b_{\rm crit}$ normalized by  
$b_{\rm B}$ for each moment is about 
$0.460$, $0.736$, $0.882$ and $0.951$, respectively.  
The angular frequency is monotonically increasing with 
respect to the impact
parameter $b$ or equivalently, the angle $\delta$, while it  
monotonically decreases with the lapse of time. The 
angular frequency at $b=0$ will vanish in the limit of 
$T_{\rm o}\rightarrow\infty$, since in this limit, the radial null geodesic  
agrees with the generator of the event horizon whose 
redshift is infinite. 

\begin{figure}
        \centerline{\epsfxsize 10cm \epsfysize 7cm \epsfbox{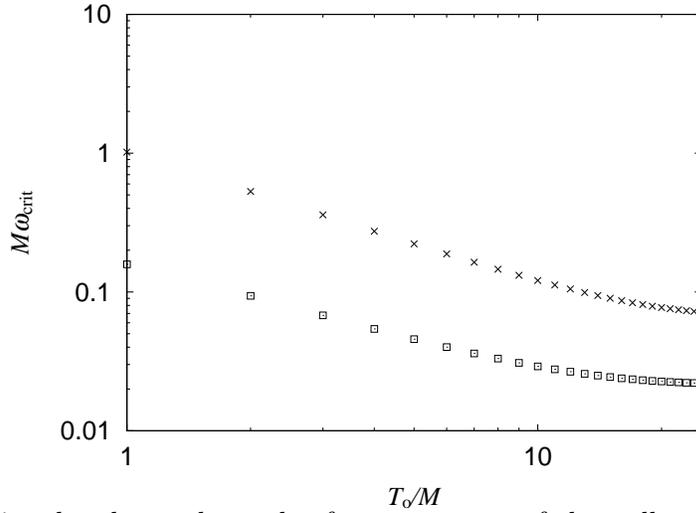}}
        \caption{We depict the observed angular frequency 
$\omega_{\rm crit}$ of the null geodesic with the largest impact parameter 
$b=b_{\rm crit}$ as a function of time $T_{\rm o}$ at $R_{\rm o}=100M$, 
where $\omega_{L}$ is set to be $10^{3}M^{-1}$. 
The crosses are the data for the the background self-similar 
dust chosen so that $W_{-}=\sqrt{2/10}$, while the squares correspond 
to the data of the background with $W_{-}=\sqrt{3/10}$. 
The observed angular frequency $\omega_{\rm crit}$ approaches 
to some finite value for $T\rightarrow \infty$. 
} 
\label{fg:omega-crit}
\end{figure}

The angular frequency $\omega_{\rm crit}$ 
is depicted as a function of the time  
$T_{\rm o}$ in Fig.\ref{fg:omega-crit}. This figure suggests  
that $\omega_{\rm crit}$ approach to some finite value in the limit 
of $T\rightarrow\infty$. Hence the naked singularity will be observed as 
a ring for sufficiently late time. 

\section{Summary and Discussion}

We studied non-radial null geodesics emanating from the central 
naked singularity formed by the gravitational collapse 
of a self-similar dust sphere. The existence of the 
non-radial singular null geodesics means that the central naked 
singularity is observed as a disk if it emits null geodesics of all 
possible impact parameters. 
The observed singular null geodesic with the largest impact parameter 
$b=b_{\rm crit}$ 
gives the angular diameter $\Delta$ of the naked singularity. 
The angular diameter $\Delta$ is time dependent; it monotonically grows 
and approaches to the value $3\sqrt{3}M/R_{\rm o}$ for 
the observer at $R=R_{\rm o}\gg M$. 
The asymptotic value of the angular diameter  
comes from the geometry of the exterior Schwarzschild region. 

As shown by Nolan and Mena\cite{ref:NM02}, in the case of the 
self-similar LTB solution, the topology of the central naked singularity 
is not uniform, being point-like on one region and spherical on another. 
In this article, we have shown that the point-like region 
of the central naked singularity is a portion from which 
singular null geodesics of the largest impact parameter 
$b=b_{\rm crit}$ emanates. In other words, the outer boundary of the 
observed disk constitutes the null geodesics from the point-like 
portion of the central naked singularity. 
On the other hand, all the other singular null geodesics emanate  
from the spherical region of the central naked singularity. 

We also studied the angular frequencies of the physical fields 
which propagate along the singular null
geodesics. It is well known fact that the redshift between this  
naked singularity and any timelike observer is infinite. 
Hence even if some physical field with a finite frequency 
is excited at the central naked singularity, it cannot carry its energy 
from the central naked singularity to the outside. From this result, 
it seems to be likely that 
no physical information comes form the central naked 
singularity to any observer. However, it might be terrible to assume that 
some physical field with a $finite$ frequency are generated at 
the naked singularity at which the energy density of the dust fluid and 
spacetime curvature are $infinite$. 
Graviton might be only one physical field which can carry the 
physical information to distant observers from the central naked 
singularity since it might not be scattered nor absorbed 
even in the situation with the extremely high energy density 
like as in the neighborhood of the central naked singularity.  
Its frequency might be order of $\sqrt{\rho}$, that is, the inverse of the 
free fall time. Since the energy density $\rho$ near the central naked 
singularity is proportional to $1/R^{2}$ along the singular 
null geodesic, the angular 
frequency of this field generated at $R=L$ are order of $1/L$. 
Hence in this article, we have considered the null geodesics 
which start at $R=L$ with the comoving 
angular frequency $\omega_{L}=1/L$ and then have investigated 
the angular frequency $\omega_{\rm o}$ observed at $R=R_{\rm o}\gg M$. 
On the ground of dimensional analysis, if the fields are excited by the
quantum gravitational effects, $L$ will be order of the Planck length. 

As a result, we found that the angular frequency is a function 
of the impact parameter $b$, or equivalently the angle $\delta$ 
between the radial direction and the null geodesic. 
At the center $\delta=0$, the angular frequency $\omega$ 
is the smallest, while it is the largest at 
the boundary $\delta=\Delta/2$. Hence, if such null geodesics 
are really emitted, the naked singularity is observed as a disk-like   
rainbow. 

In the limit of $T\rightarrow\infty$, 
the angular frequency $\omega$ at $\delta=0$ 
vanishes while it does not for $0<\delta\leq\Delta/2$. 
The singular null geodesic at $\delta=0$
in the limit of $T\rightarrow \infty$ is the generator of the 
event horizon and hence suffers the infinite gravitational redshift. 
By virtue of this gravitational redshift of the central naked singularity, 
we might observe the gravitons in a wide range of the angular frequency. 
Hence all the gravitational wave detectors (TAMA\cite{ref:TAMA}, 
LIGO\cite{ref:LIGO}, VIRGO\cite{ref:VIRGO}, GEO\cite{ref:GEO}, 
LISA\cite{ref:LISA} and DECIGO\cite{ref:DECIGO}) now in operation 
or in plan are able to detect the graviton from this naked
singularity if sufficient amount of gravitons are excited there. 

There are several researches for more  
general spherically symmetric systems and those revealed 
that non-vanishing pressure does not necessarily 
prevent the formation of the central shell focusing naked 
singularity\cite{ref:OP88,ref:OP90,ref:Harada98,ref:HIN98}. 
In these example, there might be non-radial null geodesics 
emanating from the central naked singularity. The investigation 
of these is future work. 

\vskip0.5cm
{\large{\bf Acknowledgements}}

We are grateful to T.~Harada and the colleagues in 
Department of Physics, Osaka City University for useful discussion.
KN also thanks to the workshop of DECIGO held in 
National Astronomical Observatory of Japan in May, 2002. 

\appendix

\section{}

In this appendix, we show that there is no 
singular null geodesic of ${\cal P}\rightarrow l$ for $r\rightarrow0$ in 
the self-similar LTB spacetime. 

{\it Proof}. First we assume 
${\cal P}\rightarrow l$ in the limit $r\rightarrow 0$. 
Hence, ${\cal P}$ is written in the form 
\begin{equation}
{\cal P}=l+\delta{\cal P}(r), \label{eq:P-splitted1}
\end{equation}
where $\delta{\cal P}$ satisfies
\begin{equation}
\lim_{r\rightarrow0}\delta{\cal P}=0. \label{eq:P-star-limit1}
\end{equation}
Then in the limit $r\rightarrow0$, Eq.(\ref{eq:R-equation}) becomes 
\begin{equation}
W_{0}{}^{2}=\lim_{r\rightarrow0}
{1\over3\alpha W_{0}{}^{2}}\left\{W_{0}-\sqrt{l\Lambda_{0}r^{1-\alpha}
\over2\delta{\cal P}}\right\}
\left\{W_{0}{}^{3}+2r^{{3\over2}(1-\alpha)}\right\}.
\end{equation}
Since the right hand side of the above equation should be finite, 
$\alpha$ is less than unity. Therefore, 
in the limit of $r\rightarrow 0$,  
Eq.(\ref{eq:P-equation}) leads to 
\begin{equation} 
{d\delta{\cal P}\over d\ln r}\longrightarrow {l\over3}.
\end{equation}
Integrating the above equation, we obtain 
\begin{equation}
\delta{\cal P}\longrightarrow {l\over3}\ln r ~~~~~~{\rm for}~~r\longrightarrow0.
\end{equation}
This behavior is in contradiction to the assumption 
(\ref{eq:P-star-limit1}). $\Box$

\section{}

In the self-similar LTB solution, there is a 
homothetic Killing vector $\xi^{a}$ which satisfies 
\begin{equation}
\pounds_{\xi}g_{ab}=g_{ab}.
\end{equation}
In this spacetime, its components are given by
\begin{equation}
\xi^{\mu}=(t,r,0,0). 
\end{equation}
We can easily see that $C_{\xi}\equiv -k^{\mu}\xi_{\mu}$ is constant 
along a null geodesic with tangent $k^{\mu}$. By virtue of this 
fact, we obtain an analytic expression for $\cal P$\cite{ref:JD92}. 

The self-similar variable $x$ is expressed by $W$ as
\begin{equation}
x={2\over 3\Lambda_{0}{}^{1/2}}(1-W^{3}). \label{eq:X-relation}
\end{equation}
Hence $R'$ can be also expressed by $W$ as 
\begin{equation}
R'=W^{2}+r(\partial_{r}W^{2})_{t}
={W^{3}+2\over 3W}. \label{eq:R-prime}
\end{equation} 

The conserved quantity associated with the homothetic Killing 
vector is 
\begin{equation}
C_{\xi}=tk^{t}-r{R'}^{2}k^{r}. \label{eq:conserved}
\end{equation}
From the null condition $k^{\mu}k_{\mu}=0$ and Eq.(\ref{eq:conserved}), 
we obtain for non-vanishing $C_{\xi}$, 
\begin{equation}
{\cal P}={C_{\xi}W^{2}\over x^{2}-{R'}^{2}}
\left\{x\pm R'\sqrt{1-(x^{2}-{R'}^{2})
l^{2}/(C_{\xi}W^{2})^{2}}\right\}. \label{eq:P-sol}
\end{equation}
When $C_{\xi}=0$ but $l$ does not vanish, the positive root of 
${\cal P}$ is 
\begin{equation}
{\cal P}={lR'\over \sqrt{{R'}^{2}-x^{2}}}.\label{eq:P-sol-2}
\end{equation}
For the case of $C_{\xi}=0=l$, the solution of Eq.(\ref{eq:R-eq}) 
is $W_{*}=0$. Using this solution, Eq.(\ref{eq:P-eq}) can be easily 
integrated as
\begin{equation}
{\cal P}={P_{0}\over w}.\label{eq:P-sol-3}
\end{equation}
We numerically integrate Eqs.(\ref{eq:P-eq}) and (\ref{eq:Pdagger-eq}). 
Then the solutions (\ref{eq:P-sol}), (\ref{eq:P-sol-2}) 
and (\ref{eq:P-sol-3}) are used to check the accuracy of the numerical 
integrations. In our numerical calculations, the relative error 
estimated by using these analytic solutions for $\cal P$ is 
less than $10^{-7}$.


%

\vskip0.5cm

\vskip0.5cm

\vskip1cm

\newpage

\end{document}